\documentclass[conference]{IEEEtran}
\IEEEoverridecommandlockouts
\usepackage{cite}
\usepackage{overpic}
\usepackage{amsmath,amssymb,amsfonts}
\usepackage{graphicx}
\usepackage{textcomp}
\usepackage{xcolor}
\usepackage{float}
\usepackage{amsthm}
\usepackage{graphicx}
\usepackage{epstopdf}
\usepackage{subfigure}
\usepackage{adjustbox}
\usepackage{stfloats}
\usepackage{amsmath,bm}
\usepackage{amsfonts}
\usepackage{amssymb}
\usepackage{color}
\usepackage{multirow}
\usepackage{multicol}
\usepackage{soul,xcolor}
\usepackage{algorithm}
\usepackage{algpseudocode}%

\theoremstyle{plain}

\newcommand{\vect}[1]{\mathbf{#1}}

\def\Ttran{\mbox{\tiny $\mathrm{T}$}}

\begin{document}

\title{Energy-Efficient Cell-Free Massive MIMO \\ with Wireless Fronthaul
\thanks{This work has been funded by Celtic-Next project RAI-6Green partly supported by Swedish funding agency Vinnova. The work by \"O. T. Demir was supported by 2232-B International Fellowship for Early Stage Researchers Programme funded by the Scientific and Technological Research Council of T\"urkiye. E.~Bj\"ornson was supported by the FFL18-0277 grant from SSF.}
}

\author{\IEEEauthorblockN{Ozan Alp Topal\IEEEauthorrefmark{2}, Özlem Tuğfe Demir\IEEEauthorrefmark{1}, Emil Björnson\IEEEauthorrefmark{2}, and Cicek Cavdar\IEEEauthorrefmark{2}}
\IEEEauthorblockA{ {$^\dagger$Department of Computer Science, KTH Royal Institute of Technology, Kista, Sweden
		} \\
  {$^*$Department of Electrical-Electronics Engineering, TOBB University of Economics and Technology, Ankara, Turkiye
		} \\
		\IEEEauthorblockA{E-mail: \IEEEauthorrefmark{2}\{oatopal, emilbjo, cavdar\}@kth.se, \IEEEauthorrefmark{1}ozlemtugfedemir@etu.edu.tr}
}
}

\maketitle

\begin{abstract}
Cell-free massive MIMO improves the fairness among the user equipments (UEs) in the network by distributing many cooperating access points (APs) around the region while connecting them to a centralized cloud-computing unit that coordinates joint transmission/reception. However, the fiber cable deployment for the fronthaul transport network and activating all available antennas at each AP lead to increased deployment cost and power consumption for fronthaul signaling and processing.  To overcome these challenges, in this work, we consider wireless fronthaul connections and propose a joint antenna activation and power allocation algorithm to minimize the end-to-end (from radio to cloud) power while satisfying the quality-of-service requirements of the UEs under wireless fronthaul capacity limitations. 
The results demonstrate that the proposed methodology of deactivating antennas at each AP reduces the power consumption by $50\%$ and $84\%$ compared to the benchmarks based on shutting down APs and minimizing only the transmit power, respectively. 
\end{abstract}

\vspace{-3mm}



%
\section{Introduction}

Cell-free massive MIMO (multiple-input multiple-output) is a promising candidate for future mobile technologies by improving the fairness among the user equipments (UEs) in the network. This benefit is provided by distributing many access points (APs) that are capable of joint transmission/reception in a region. This cooperation is satisfied by tight synchronization and by connecting them to a centralized cloud unit, responsible for some physical layer (PHY) or higher layer operations. Separating network functions between the centralized (or distributed) cloud processing and radio units is an active research area, where the energy saving by this centralization is studied in centralized radio access network (C-RAN), cloud-RAN \cite{cloudRAN}, and more recently Open-RAN literature \cite{ORAN}.   

In the early works of the cell-free massive MIMO network, the connection to the centralized unit is assumed to be carried out by an optical transport network, requiring fiber cable deployment for all distributed APs \cite{cell-free-book}. While this way of deployment for cell-free massive MIMO networks provides robust and fair performance, the network power consumption and deployment cost become considerably high, limiting the potential realization of cell-free networks \cite{ozlem_jsac}. In \cite{riera2023energy}, the energy saving by efficient carrier/symbol shutdown and UE-AP association has been investigated, without considering the end-to-end power consumption resulting from the cloud   processing.  In \cite{ozlem_jsac}, end-to-end power consumption performance of network functional splits has been studied for cell-free massive MIMO, where the end-to-end power minimization approach has been shown to reduce power consumption up to $30\%$ compared to the case where radio resources are optimized
independently from the cloud. However, the authors consider fiber cable connections between APs and the cloud unit, increasing the deployment cost of the APs, and hindering the real-life applicability of the large-scale networks. The performance of wireless fronthaul for cell-free networks has been analyzed due to the low-cost implementation benefits compared to optical fronthaul deployment \cite{umurhan, wireless_f}; however, the energy-efficient operation has not been addressed.  

In this work, we propose a joint power allocation and antenna activation scheme for an end-to-end energy-efficient cell-free massive MIMO network with wireless fronthaul. We first propose a novel wireless fronthaul access scheme for the APs based on a combination of time-division multiple access (TDMA) and space-division multiple access (SDMA). This approach groups APs into SDMA clusters and maximizes the orthogonality between their channels. Then, we formulate and solve the joint antenna activation and power allocation problem to minimize end-to-end network energy consumption guaranteeing quality-of-service (QoS) requirements of the UEs by utilizing a novel block-coordinate descent algorithm. The proposed algorithm provides load balancing both over wireless fronthaul and radio access, which has not been considered before for cell-free massive MIMO. The numerical results demonstrate that activation/deactivation of antennas at APs reduces the power consumption by more than $50\%$ compared to  controlling only the activation/deactivation of APs, thanks to the finer granularity of the control proposed by this work.  

\begin{figure}[tb!]
    \centering
    \includegraphics[width=0.8\linewidth]{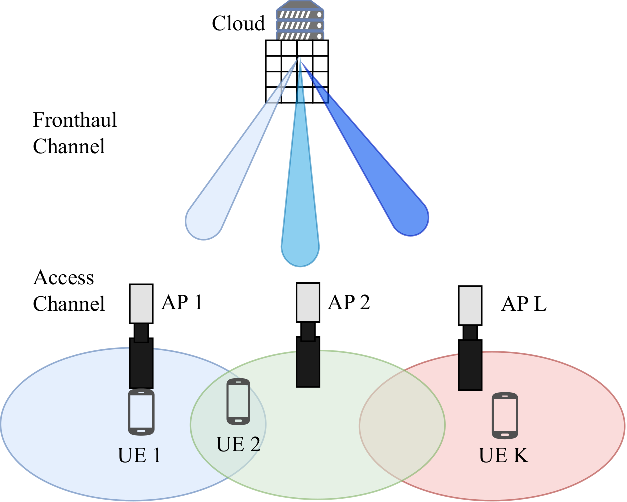}
     \vspace{-5mm}
    \caption{Cell-free massive MIMO architecture with wireless fronthaul.   
    }
    \vspace{-5mm}
    \label{fig:system_model}
\end{figure} 

\section{System Model}

We consider the downlink of a cell-free massive MIMO system operating in time-division duplex (TDD) mode as illustrated in Fig.~\ref{fig:system_model}. The system consists of $L$ APs connected to the cloud via a high-frequency (mmWave) wireless fronthaul link, where APs and the cloud have line-of-sight (LOS) connectivity.  The APs serve $K$ single antenna UEs over a mid-band frequency (sub-6 GHz) channel. The channel between APs and the cloud unit will be referred to as the fronthaul channel, while the channel between APs and UEs will be referred to as the access channel. For the access links, we assume uncorrelated Rayleigh fading channels as in \cite{interdonato2020local}. Each AP is equipped with $M^\mathrm{ac}$ antennas for the access channel and $M^\mathrm{frh}$ antennas for the fronthaul channel. In the proposed system, we consider that each AP can activate or deactivate its access antennas based on the QoS requirements, fronthaul load limitations, and energy minimization. Therefore, we denote the activated antennas at AP $l$  by $M_l \in \{0,\ldots,M^\mathrm{ac}\}$, where $M_l=0$ means that AP is deactivated. The bandwidth utilized in the access channel, and in the fronthaul channel are denoted as $B^{\mathrm{ac}}$ and $B^{\mathrm{frh}}$, respectively. 

We let $\tau_c$ denote the number of symbols in a TDD frame, where it consists of uplink training (with $\tau_p$ symbols for the training) and downlink data transmission (with $\tau_d=\tau_c-\tau_p$ symbols). We follow the same uplink training phase as described in \cite{interdonato2020local}, and due to the space limitation we omit the explanations from this paper. 

\subsection{Downlink Data Transmission in Access Channel}

 In this work, we consider functional split Option $7.2$, where all PHY operations higher than precoding are done in the centralized cloud, where precoding and lower operations are done in the APs, corresponding to the distributed precoding schemes in cell-free massive MIMO literature \cite{cell-free-book}. After receiving the downlink message signals from the cloud, APs can simultaneously apply precoding and transmit the same message signal to enable a coherently enhanced signal at each receiving UE. The transmit signal from AP $l$ can be given as 
\begin{equation}
    \vect{x}_l = \sum_{k=1}^K \sqrt{\rho_{l,k}} \vect{w}_{l,k} \varsigma_{k},
\end{equation}
where $\rho_{l,k}$ is the  transmit power assigned for UE $k$ at AP $l$. $\vect{w}_{l,k} \in \mathbb{C}^{M_l}$ is the precoding vector used by AP $l$ towards UE $k$ with $\mathbb{E}\left\{\|\vect{w}_{l,k} \|^2\right\} = 1$. $\varsigma_{k}$ is the unit-power downlink data symbol of UE $k$,  $\mathbb{E}\left\{|\varsigma_{k}|^2\right\}=1$. 

We consider local protective partial zero-forcing (PPZF) as the downlink precoding \cite{interdonato2020local}. Zero-forcing provides interference cancellation, but it significantly reduces the array gain. On the other hand, maximum ratio transmission (MRT) (or conjugate beamforming) maximizes the array gain, while resulting in high interference. PPZF is capable of providing the right balance between the array gain and interference cancellation compared to other distributed precoding schemes. In PPZF, each AP divides UEs into two distinct sets: strong-channel UEs, and weak-channel UEs. Then, APs utilize ZF precoding for the strong-channel UEs, and protective MRT for the weak-channel UEs. The protectiveness of MRT comes from canceling out the interference of weak-channel UEs to strong-channel UEs, creating protection for the strong-channel UEs. We let $\mathcal{S}_l$ and $\mathcal{W}_l$ denote the sets of strong-channel UEs and weak-channel UEs at AP $l$, respectively, where $\mathcal{S}_l \bigcap \mathcal{W}_l = \varnothing$.  

\begin{figure*}[tb]
\begin{equation}
    \operatorname{SINR}_k =\frac{\left(\sum_{l=1}^L \sqrt{\left(M_l-\tau_{\mathcal{S}_l}\right) \rho_{l, k} \gamma_{l, k}}\right)^2}{\sum_{t \in \mathcal{P}_k \backslash\{k\}}\left(\sum_{l=1}^L \sqrt{\left(M_l-\tau_{\mathcal{S}_l}\right) \rho_{l, t} \gamma_{l, k}}\right)^2+\sum_{t=1}^K \sum_{l=1}^L \rho_{l, t}\left(\beta_{l, k}-\delta_{l, k} \gamma_{l, k}\right)+\sigma^2}.
    \label{eq:SINR}
\end{equation}
\hrulefill
\vspace{-4mm}
\end{figure*}

The exact expressions of precoding vectors are also omitted due to space limitation, but the readers can refer to \cite{interdonato2020local}. A lower bound on the spectral efficiency (SE) for UE $k$ is given by $\mathrm{SE}_k = \frac{\tau_c - \tau_p}{\tau_c} \log_2(1+ \mathrm{SINR}_k)$, where $\mathrm{SINR}_k$ is named as the effective signal-to-interference-plus-noise ratio (SINR) of UE $k$, and for the considered precoding scheme can be given as in \eqref{eq:SINR} as shown at the top of the next page, where $\sigma^2$ is the receiver noise variance. Here, $\beta_{l,k}$ and $\gamma_{l,k}$ represent the large-scale fading coefficient for the channel from AP $l$ to UE $k$ and the mean-square of the corresponding channel estimate, respectively. $\delta_{l,k}$ denotes the membership decision of UE $k$, where 
\begin{equation}
    \delta_{l,k} = \begin{cases}
    & 1, \quad \text{if } k \in \mathcal{S}_l, \\
    & 0, \quad \text{if } k \in \mathcal{W}_l.
    \end{cases}
\end{equation}$\tau_{\mathcal{S}_l} \leq \tau_{p}$ denotes the number of pilot signals for the strong-channel UEs at AP $l$.
The steps to derive the SINR expression are skipped due to the page limitation; however, the proof can be obtained by following the steps given in Appendix C of \cite{interdonato2020local} by assuming that each AP can have a different number of antennas.

\subsection{GOPS Analysis}
3GGP defines several split options that allow the network to carry out some of the PHY functions in the cloud while some are in the APs. In this work, we consider split Option 7.2, where the radio frequency (RF) layer and lower PHY  operations are carried out at the AP, while higher PHY processes such as modulation, and coding are carried out at the cloud. We will keep the discussion on the functional split selection to the extension of this work. The giga operations per second (GOPS) for the operations considered in this work are given in Table \ref{tab:GOPS_table} \cite{Debaillie2015a,malkowsky2017world,desset2016massive,ozlem_jsac}. $\mathrm{W}_r$ and $\mathrm{SE}_r$ denote the ratio of the bandwidth and the ratio of the SE of a UE for this work to the reference setup \cite{Debaillie2015a}. In the reference setup, $20$\,MHz bandwidth is chosen, and the SE is equal to $6$\,bit/s/Hz. The binary $r_{i,l}$ takes the value of $1$ if AP $l$ serves UE $i$ and zero otherwise. $T_s$ is the OFDM symbol duration, $N_{\rm DFT}$ is the DFT size, and $N_{\rm used}$  is the number of used subcarriers. The GOPS at AP $l$ can be calculated as
\begin{equation}
   C_{\mathrm{AP},l} = C_{\mathrm{filter},l}  + C_{\mathrm{DFT},l} +  C_{\mathrm{map},l} + C_{\mathrm{prec},l}.   
\end{equation}
The GOPS at the cloud general-purpose processor (GPP) is
\begin{equation}
   C_{\mathrm{GPP}} = \sum_{l=1}^L \left(C_{\mathrm{mod}, l}  + C_{\mathrm{coding}, l} +  C_{\mathrm{network}, l} \right).   
\end{equation}

\begin{table}[tb]
\vspace{-4mm}
     \centering
     \caption{GOPS calculations for different functions.}
     \vspace{-2mm}
     \begin{tabular}{l|l c}
     Function & GOPS Expression \\ \hline
     $C_{\mathrm{filter}, l}$ & ${40 M_l f_s}/{10^9}$ \\ 
     $ C_{\mathrm{DFT}, l}$ & $\frac{8 M_l N_{\mathrm{DFT}} \log_2(N_{\mathrm{DFT}})}{T_s10^9}$ \\ 
       $C_{\mathrm{map},l}$   &  $ 1.3 \mathrm{W}_r \mathrm{SE}^{1.5}_r   \sum_{i=1}^K r_{i, l} $   \\
          $ C_{\mathrm{prec}, l}$ & $\left(\frac{8 M_l \tau_d N_{\mathrm{used}}}{T_s 10^9 \tau_c} \right)  \sum_{i=1}^K r_{i, l}$ \\
    \hline     \hline
       $C_{\mathrm{mod},l}$   & $ 1.3 \mathrm{W}_r M_l  $ \\
      $C_{\mathrm{coding},l}$ &  $  5.2 \mathrm{W}_r \mathrm{SE}_r  \sum_{i=1}^K r_{i, l}$ \\
      $C_{\mathrm{network},l}$ & $ 8 \mathrm{W}_r \mathrm{SE}_r  $ \\
      
     \end{tabular}
     \label{tab:GOPS_table}
     \vspace{-0.6cm}
 \end{table}

\subsection{End-to-End Power Consumption Model}
We consider both AP-side and cloud-side power consumption for accurate characteristics of the network as in  \cite{ozlem_jsac}. Power consumption by APs can be calculated by 
\begin{equation}
  P_{l} =  \sum_{l=1}^L M_l P_{\mathrm{st}}  + \Delta^{\rm tr} \sum_{k=1}^{K} \rho_{l,k} + P_{\mathrm{proc},l}, 
\end{equation}
where $M_l P_{\mathrm{st}}$ is the hardware-dependent static power consumption at AP $l$, $\sum_{k=1}^{K} \rho_{l,k}$ is the load-dependent total transmit power with the slope $\Delta^{\rm tr}$ at AP $l$, and $P_{\mathrm{proc},l}$ is the power consumption by the processing done at AP $l$. The latter is given as 
\begin{equation}
 P_{\mathrm{proc},l} = P^{\mathrm{proc}}_{0} + \Delta^{\mathrm{proc}}_{\mathrm{AP}} \frac{C_{\mathrm{AP},l}}{C_{\mathrm{AP, max}}},
\end{equation}
where $P^{\mathrm{proc}}_{0}$ is the idle processing power, $C_{\rm AP,max}$ is the total GOPS capability of AP $l$, and $\Delta_{\rm AP}^{\rm proc}$ is the slope of the corresponding load-dependent power consumption.
On the cloud side, the power consumption can be given as
\begin{align}
P_{\mathrm{Cloud }}&=  P_{\mathrm{fixed}}+ \Delta^{\rm tr}\sum_{l=1}^L \bar{p}_l \nonumber\\
&\quad +
\frac{1}{\sigma_{\mathrm{cool }}}\left( P_{\mathrm{comp}} + \Delta_{\mathrm{GPP}}^{\mathrm{proc}} \frac{C_{\mathrm{GPP}}}{C_{\mathrm{GPP}}^{\max }}\right),
\end{align}
where $\bar{p}_l$ is the transmit power utilized for fronthaul link between the cloud and AP $l$. $P_{\rm fixed}$ is the load-independent fixed power consumption, $P_{\rm comp}$ is the idle processing power, and other parameters are defined in the same way.

The end-to-end power consumption can be expressed as 
\begin{align}
      & P_{\mathrm{tot}} = \bar{P}_{\mathrm{fixed}} + c_0 \sum_{k=1}^{K} \sum_{l=1}^{L} \rho_{l,k} + c_1 \sum_{l=1}^{L}  M_l + c_2 \sum_{l=1}^{L} \mathbb{I}(M_l) \nonumber\\ & + c_3\sum_{l=1}^{L}  \sum_{k=1}^{K} \mathbb{I}(\rho_{l,k}) + c_4 \sum_{l=1}^{L} M_l \left( \sum_{k=1}^{K} \mathbb{I}(\rho_{l,k}) \right)
      + c_5 \sum_{l=1}^{L} \bar{p}_l, 
\end{align}
where $\mathbb{I}(\cdot)$ is the indicator function, which is equal to one if the input of the function is greater than zero, and equal to zero otherwise. $\bar{P}_{\mathrm{fixed}} = {P}_{\mathrm{fixed}} + \frac{P_{\mathrm{comp}}}{\sigma_{\mathrm{cool }}}$ is the fixed power consumption that is ignored in the optimization problem, and later included in simulation results. The coefficients can be obtained as $c_0 = \Delta^{\rm tr}$, $c_1 = P_{\mathrm{st}} + 1.3 \mathrm{W}_r \frac{\Delta_{\mathrm{GPP}}^{\mathrm{proc}} }{C_{\mathrm{GPP}}^{\max }\sigma_{\mathrm{cool }}} + \frac{\Delta^{\mathrm{proc}}_{\mathrm{AP}} }{C_{\mathrm{AP, max}}} \left[ \frac{40 f_s}{10^9} + \frac{8 N_{\mathrm{DFT}} \log_2(N_{\mathrm{DFT}})}{T_s10^9}  \right] $, $c_2 = \frac{\Delta_{\mathrm{GPP}}^{\mathrm{proc}} }{C_{\mathrm{GPP}}^{\max }\sigma_{\mathrm{cool }}}\left(8 \mathrm{W}_r \mathrm{SE}_r\right) + P^{\mathrm{proc}}_{0}$, $c_3 = \frac{\Delta^{\mathrm{proc}}_{\mathrm{AP}} }{C_{\mathrm{AP, max}}} \left( 1.3 \mathrm{W}_r \mathrm{SE}^{1.5}_r  \right)  + \frac{\Delta_{\mathrm{GPP}}^{\mathrm{proc}} }{C_{\mathrm{GPP}}^{\max }\sigma_{\mathrm{cool }}} \left(5.2 \mathrm{W}_r \mathrm{SE}_r\right) $, $c_4 = \frac{\Delta^{\mathrm{proc}}_{\mathrm{AP}} }{C_{\mathrm{AP, max}}} \left(\frac{8 \tau_d N_{\mathrm{used}}}{T_s 10^9 \tau_c} \right)  $, $c_5 =  \Delta^{\rm tr}$.

\section{Wireless Fronthaul Access}
We consider a combination of TDMA/SDMA for the fronthaul channel.  The cloud unit is equipped with $M_c$ antennas driven by  $N_c$ RF chains that $N_c \ll M_c$. The cloud divides APs into distinct groups with a maximum size of  $N_c$. Then, the TDMA protocol is applied between groups.

\subsection{Fronthaul Signal Transmission}
We let $\mathcal{L}_i$ denote the $i$th group of APs, where $\sum_{i=1}^{\lceil L/N_c \rceil}|\mathcal{L}_i| = L$. The received signal at AP $l$ in $\mathcal{L}_i$ is 
\begin{equation}
    \vect{y}_{l} = \vect{G}_{l} \vect{F}_i \vect{W}_i \vect{s}_i + \vect{n}_{l},
\end{equation}
where $ \vect{G}_{l} \in \mathbb{C}^{   M^{\mathrm{frh}} \times M_c }$ is the downlink fronthaul channel between the cloud and AP $l$. Since the fronthaul links are assumed in LOS, and the AP deployments are static, the fronthaul channel is assumed to be perfectly known in the cloud.  

$\vect{s}_i \in \mathbb{C}^{|\mathcal{L}_i|}$ denotes the downlink message signal of APs in $\mathcal{L}_i$. 
$\vect{F}_i \in \mathbb{C}^{M_c \times N_c}$ and $\vect{W}_i\in \mathbb{C}^{N_c \times |\mathcal{L}_i|}$ are the analog and digital precoding matrices. The APs also perform analog combining, after which the corresponding received signal can be represented as 
\begin{equation}
    {\hat{y}}_{l} = \vect{v}_l^H \vect{G}_{l} \vect{F}_i \vect{W}_i \vect{s}_i + \vect{v}_l^H \vect{n}_{l},
\end{equation}
where $\vect{v}_l \in \mathbb{C}^{M^{\mathrm{frh}} }$. We assume that the cloud chooses the columns of the $\vect{F}_i$ as the array response vectors in the directions of the corresponding APs. Similarly, combining vectors are chosen in the angle of arrival from the cloud to the corresponding AP. 
By including the effects of analog beamforming into the channel, we can characterize equivalent channel representation as $(\vect{g}^{\mathrm{eq}}_l)^H = \vect{v}_l^H \vect{G}_{l} \vect{F}_i $, and $\vect{\Bar{G}}_i = [\vect{g}^{\mathrm{eq}}_1, \ldots, \vect{g}^{\mathrm{eq}}_{|\mathcal{L}_i|}]^H$. Applying zero-forcing precoding at the cloud, we obtain the achievable data rate of AP $l$ as \cite[Ch. 6]{bjornson2024introduction}
\begin{equation}
R^f_l = t_iB^{\mathrm{frh}}\log_2 \left(1+ \Lambda_{ll} \bar{p}_l \right),
\end{equation}
where $\Lambda_{ll}$ is equal to $1/(\sigma^2\left[(\vect{\Bar{G}}_i\vect{\Bar{G}}_i^H)^{-1}\right]_{l,l})$ and $t_i$ is the allocated time portion to group $i$, where $\sum_{i=1}^{I} t_i = 1$. Here, $I$ is defined as the number of groups, i.e., $I=\lceil L/N_c \rceil$.

\subsection{AP Grouping for Fronthaul Access}
In AP grouping, we aim to maximize the orthogonality of the channels in a group to reduce the possible interferences.  We utilized chordal distance as the main metric to model the orthogonality between channels. The chordal distance between fronthaul channels of AP $l$ and AP $l'$  is defined as 
$\zeta_{l,l'} = \frac{\left|{\vect{g}^{\mathrm{eq}}_{l}}^H\vect{g}^{\mathrm{eq}}_{l'}\right|}{\| \vect{g}^{\mathrm{eq}}_{l} \| \| \vect{g}^{\mathrm{eq}}_{l'} \| },$
where $\zeta_{l,l'} \in [0,1]$. Grouping APs with lower chordal distance corresponds to grouping APs with higher orthogonality. As a result, possible interference among APs is minimized.

One way to group APs can be to minimize the maximum sum chordal distance of a group. We let $\alpha_{l,i} \in \{0,1\}$ to denote the membership of AP $l$ in group $i$, and $\boldsymbol{\alpha}_i = [\alpha_{1,i}, \ldots, \alpha_{L,i}]^T$. We also concatanate all chordal distances into a matrix, $\boldsymbol{\zeta} \in \mathbb{R}^{L\times L}$, where $[\boldsymbol{\zeta}]_{l,l'} = \zeta_{l,l'}$ and diagonal entries being zero. We define a  binary matrix, $\vect{A}_i = \boldsymbol{\alpha}_i\boldsymbol{\alpha}^{\Ttran}_i \in \left\{0, 1\right\}^{L \times L}$ and $[\vect{A}_i]_{l,l'}= a_{l,l',i}= \alpha_{l,i} \alpha_{l',i}$.  The elements of the matrix can be replaced by the following constraints:
\begin{equation}
\begin{aligned}
&a_{l,l',i} \leq \alpha_{l,i}, \quad a_{l,l',i} \leq \alpha_{l',i}, \quad
 a_{l,l',i} \geq \alpha_{l,i}+\alpha_{l',i}-1,
\label{eq:constraint:binary_matrix}
\end{aligned}
\end{equation} 
where $a_{l,l',i}  \in \{0,1\}$. The optimization problem is given as
\begin{subequations} \label{eq:group_opt1:problem}
\begin{align}
 & \underset{\{\alpha_{l,i}, a_{l,l',i} \},\varsigma}{\text{minimize}}  \quad \varsigma \label{eq:group_opt1:objective} \\ & \textrm{subject to} \quad \eqref{eq:constraint:binary_matrix} \quad \nonumber \\ &
 \operatorname{Tr}( \boldsymbol{\zeta}\vect{A}_i)
 \leq \varsigma, \quad \forall i
 \\ &
 \sum_{i=1}^{I} \alpha_{l,i} = 1, \quad \forall l, \quad \sum_{l=1}^{L} \alpha_{l,i} \leq N_c, \quad \forall i  \label{eq:group_opt:limitgroupsize} \\ &
 \alpha_{l,i}, a_{l,l',i} \in \{0,1\}, \quad \forall l,l',i \label{eq:group_opt1:binary}.
\end{align}
\end{subequations}

Global optimum can be obtained for this problem by using a branch-and-bound algorithm which in this work, it is implemented by MOSEK with CVX in MATLAB.

\section{End-to-End Power Minimization}
The main purpose of this paper is to minimize the end-to-end power consumption of the cell-free massive MIMO network. This problem can be formulated as

\begin{subequations} \label{eq:power_min_opt0:problem}
\begin{align}
 & \underset{\{M_l, \rho_{l,k}, \bar{p}_l, t_i \}}{\text{minimize}} \quad  c_0 \sum_{k=1}^{K} \sum_{l=1}^{L} \rho_{l,k} + c_1 \sum_{l=1}^{L}  M_l + c_2 \sum_{l=1}^{L} \mathbb{I}(M_l) \nonumber\\ & + c_3\sum_{l=1}^{L}  \sum_{k=1}^{K} \mathbb{I}({\rho}_{l,k}) + c_4 \sum_{l=1}^{L} M_l \left( \sum_{k=1}^{K} \mathbb{I}({\rho}_{l,k}) \right)
      + c_5 \sum_{l=1}^{L} \bar{p}_l \label{eq:power_min_opt0:objective} \\
     & \textrm{subject to} \nonumber \\ &\operatorname{SINR}_k\geq \upsilon_k, \quad \forall k \label{eq:power_min_opt0:SINR_constraint} \\
 &  t_i B^{\mathrm{frh}} \log_2\left( 1 + {\Lambda_{ll} \bar{p}_l}  \right) \geq    O_{7.2}\sum_{k=1}^{K} \mathbb{I}({\rho}_{l,k}), \quad \forall l \label{eq:power_min_opt0:wireless_fronthaul_rate} 
 \\&
\sum_{l = 1}^{L} \alpha_{l,i} \bar{p}_l \leq P_f, \quad  \forall i  \label{eq:power_min_opt0:wireless_fronthaul_power} 
   \\&
\sum_{i=1}^{I} t_i \leq 1 \label{eq:power_min_opt0:TDMA}  \end{align}
      \begin{align}& 
 \sum_{k=1}^{K} \rho_{l,k} \leq  P_t \quad \forall l\label{eq:power_min_opt0:access_power_limit} \\&
M_l \in \{\tau_{\mathcal{S}_l} +1,\ldots,M^{\mathrm{ac}}\}, \quad \forall l \label{eq:power_min_opt0:integer} .
\end{align}
\end{subequations}
The objective function, \eqref{eq:power_min_opt0:objective}, is the end-to-end total power consumption when the fixed power component is neglected since it does not change with the optimization variables. \eqref{eq:power_min_opt0:SINR_constraint} ensures that effective SINR at UE $k$ is higher than equal to the threshold value $\upsilon_k$. \eqref{eq:power_min_opt0:wireless_fronthaul_rate} ensures that the fronthaul rate for AP $l$ is higher than or equal to the required fronthaul rate for the considered split Option $7.2$, where $O_{7.2} =\frac{2N_{\rm used}N_{\mathrm{bits}}}{T_s} $ with $N_{\mathrm{bits}}$ being the number of bits to quantize the signal samples. \eqref{eq:power_min_opt0:wireless_fronthaul_power} and \eqref{eq:power_min_opt0:access_power_limit} limit the transmit power in the fronthaul and access links, respectively. \eqref{eq:power_min_opt0:TDMA} is the time allocation limit for the TDMA part of the fronthaul channel. \eqref{eq:power_min_opt0:integer} ensures that the number of active antennas at AP $l$, $M_l$ is an integer variable smaller than or equal to the deployed number of antennas at AP $l$. \eqref{eq:power_min_opt0:problem} is a non-convex problem with a combinatorial nature due to the integer variables and indicator functions.

We first relax $M_l$ to a continuous variable  $\hat{M}_l$. For mathematical convenience, we replace  \eqref{eq:power_min_opt0:integer} with $0 \leq \tilde{M}_l \leq  M^{\mathrm{ac}} - \tau_{\mathcal{S}_l}$, where $\tilde{M}_l = \hat{M}_l - \tau_{\mathcal{S}_l}$. Before handling the indicator functions, we reformulate the SINR constraints  by utilizing the auxiliary variables $z_{l,k} = \sqrt{\tilde{M}_l \rho_{l,k}}$, and $\vect{z}_k = [ z_{1,k}, \ldots, z_{L,k}]^T$,  $ \boldsymbol{\bar{\gamma}}_k = [ \sqrt{\gamma_{1,k}}, \ldots, \sqrt{\gamma_{L,k}}]^T$.  The reformulation of \eqref{eq:power_min_opt0:SINR_constraint} can be given by 
\begin{equation}
    \left \| \begin{bmatrix} &\sqrt{\upsilon_k} \iota_{1,k}\boldsymbol{\bar{\gamma}}_k^T \vect{z}_1 \\ & \vdots\\ &\sqrt{\upsilon_k}  \iota_{K,k} \boldsymbol{\bar{\gamma}}_k^T \vect{z}_K \\ & \sqrt{\upsilon_k}  (\boldsymbol{\psi}_k \odot \boldsymbol{\bar{\rho}}) \\ &  \sqrt{\upsilon_k\sigma^2} \end{bmatrix} \right \| \leq \boldsymbol{\bar{\gamma}}_k^T \vect{z}_k, \quad \forall k,
    \label{eq:SOC:SINR}
\end{equation}
where $\boldsymbol{\psi}_k = [\psi_{1,k}\vect{1}_K^T \ldots,  \psi_{L,k}\vect{1}_K^T]^T$, $\psi_{l,k} = \sqrt{\beta_{l,k} - \delta_{l,k} \gamma_{l,k}}$, $\boldsymbol{\bar{\rho}} = [\bar{\rho}_{1,1}, \ldots, \bar{\rho}_{L,K}]^T$, and $\bar{\rho}^2_{l,k} = \rho_{l,k}$. $\odot$ denotes Hadamard product. $\iota_{k',k}=1$ if UE $k'$ is in $\mathcal{P}_k \backslash\{k\}$, otherwise it is equal to zero. This is a second-order cone constraint in a convex form. To guarantee the transformation of $z_{l,k}$ we introduce 
\begin{equation}
   0  \leq z_{l,k} \leq \sqrt{ \tilde{M}_l } \bar{\rho}_{l,k}, \quad \forall l,k.
      \label{eq:z_upper_bound}
\end{equation}

We  replace $\mathbb{I}(M_l)$ with a binary variable $m_l\in \{0,1\}$, where $m_l=1$ if $M_l >0$, zero otherwise. Similarly,  $\mathbb{I}(\rho_{l,k})$ can be replaced by $r_{l,k}\in \{0,1\}, \forall l, k$. The problem becomes
\begin{subequations} \label{eq:power_min_opt1:problem}
\begin{align}
 & \underset{\{\tilde{M}_l, \bar{\rho}_{l,k}, \bar{p}_l, t_i, \vect{z}_k, r_{l,k}, m_l \}}{\text{minimize}} \!\!\!\!  c_0 \sum_{l=1}^{L} \sum_{k=1}^{K} \bar{\rho}^2_{l,k} + c_1 \sum_{l=1}^{L}  \tilde{M}_l + c_2 \sum_{l=1}^{L} m_l \nonumber\\ & + c_3\sum_{l=1}^{L}  \sum_{k=1}^{K} r_{l,k} + c_4 \sum_{l=1}^{L} \tilde{M}_l \left( \sum_{k=1}^{K} r_{l,k} \right)
      + c_5 \sum_{l=1}^{L} \bar{p}_l \label{eq:power_min_opt1:objective} \\ & \textrm{subject to}  \quad \eqref{eq:power_min_opt0:wireless_fronthaul_power}, \eqref{eq:power_min_opt0:TDMA}, \eqref{eq:SOC:SINR}, \eqref{eq:z_upper_bound} \nonumber\\
      &\bar{\rho}_{l,k} \leq  r_{l,k} \sqrt{P_t}, \quad  \forall l,k 
      \end{align}
      \begin{align} & B^{\mathrm{frh}} \log_2\left( 1 + {\Lambda_{ll} \bar{p}_l}  \right) \geq   O_{7.2}\left(\frac{\sum_{k=1}^K r^2_{l,k}}{ t_i}\right) , \quad \forall l \label{eq:power_min_opt1:wireless_fronthaul_rate}  \\& 
\sum_{k=1}^{K} \bar{\rho}^2_{l,k} \leq  P_t \quad \forall l \label{eq:power_min_opt1:access_power_limit} \\&
m_{l} \leq \tilde{M}_l \leq m_l M^{\mathrm{ac}}, \quad \forall l \label{eq:power_min_opt1:M_interval} \\&
\sum_{k=1}^K r_{l,k} \leq m_l K, \quad \forall l \label{eq:power_min_opt1:AP_activation_UE_assoc} \\ &
r_{l,k}, m_l \in \{0,1\},  \quad \ \forall l,k. \label{eq:power_min_opt1:binary_variables}
\end{align}
\end{subequations}
This problem is non-convex due to $ \tilde{M}_l \left( \sum_{k=1}^{K} r_{l,k} \right)$, the binary constraints of \eqref{eq:power_min_opt1:binary_variables}, and the constraint of \eqref{eq:z_upper_bound}. 
The global optimum for this problem cannot be guaranteed, but an efficient solution can be obtained by adding auxiliary variables representing the continuous relaxation of the binary variables, separating binary and continuous variables into different sub-problems, and alternating between these sub-problems.  We first define continuous auxiliary variables, $0 \leq \tilde{r}_{l,k}, \tilde{m}_l \leq 1, \forall l,k$, which will replace the binary variables, $r_{l,k}, m_l$,  in the constraints. We also define auxiliary variables for power coefficients, $v_{l,k}$ and $u_{l,k}$, which are used in  removing nonconvexities in \eqref{eq:z_upper_bound}. Ideally, we want a final solution to satisfy $\tilde{r}_{l,k} = {r}_{l,k}$,  $\tilde{m}_{l,k} = {m}_{l,k}$, ${u}_{l,k} = {\bar{\rho}}^2_{l,k}$, ${v}_{l,k} = {\bar{\rho}}_{l,k}$. Therefore, we add a mean-square-error (MSE) penalty to minimize the error that can be caused due to the relaxations.

The first problem with the continuous variables except the auxiliary variables for power coefficients can be written as
\begin{subequations} \label{eq:power_min_opt_subprob1:problem}
\begin{align}
 & \underset{\{ \tilde{M}_l,  \bar{\rho}_{l,k}, \bar{p}_l, t_i, \vect{z}_k, \tilde{r}_{l,k}, \tilde{m}_l, u_{l,k} \}}{\text{minimize}}   c_0 \sum_{l=1}^{L} \sum_{k=1}^{K}  \bar{\rho}^2_{l,k} \nonumber\\
 &+ c_1 \sum_{l=1}^{L}  \tilde{M}_l 
   + c_4 \sum_{l=1}^{L} \tilde{M}_l \left( \sum_{k=1}^{K} r_{l,k} \right)  + c_5 \sum_{l=1}^{L} \bar{p}_l \nonumber\\ & + \lambda_1 \sum_{l=1}^{L}  \sum_{k=1}^{K} (r_{l,k} - \tilde{r}_{l,k})^2 +  \lambda_2 \sum_{l=1}^{L} (m_{l} - \tilde{m}_{l})^2\nonumber\\ &  + \lambda_3 \sum_{l=1}^{L} \sum_{k=1}^{K} (u_{l,k} - v^2_{l,k})^2 + \lambda_4 \sum_{l=1}^{L} \sum_{k=1}^{K} (\bar{\rho}_{l,k} - v_{l,k})^2
 \label{eq:power_min_subprob1:objective} \\
 & \textrm{subject to} \quad  \eqref{eq:power_min_opt0:wireless_fronthaul_power}, \eqref{eq:power_min_opt0:TDMA}, \eqref{eq:SOC:SINR}, \eqref{eq:power_min_opt1:access_power_limit} \nonumber \\
\\ & \left\| \left[ \sqrt{2} z_{l,k}, \tilde{M}_l, u_{l,k} \right] \right\| \leq \tilde{M}_l + 
u_{l,k}, \quad \forall l,k
\label{eq:power_min_subprob1:SOC_antenna_rho} \\
 & B^{\mathrm{frh}} \log_2\left( 1 + {\Lambda_{ll} \bar{p}_l}  \right) \geq   O_{7.2}\left(\frac{\sum_{k=1}^K \tilde{r}^2_{l,k}}{ t_i}\right) , \quad \forall l \label{eq:power_min_subprob1:wireless_fronthaul_rate}  \\&
\tilde{m}_l \leq \tilde{M}_l \leq \tilde{m}_l M^{\mathrm{ac}}, \quad \forall l \label{eq:power_min_subprob1:M_interval} \\   &
\sum_{k=1}^K \tilde{r}_{l,k} \leq \tilde{m}_l K, \quad \forall l \label{eq:power_min_subprob1:AP_activation_UE_assoc} \\&
u_{l,k} \leq  \tilde{r}_{l,k} P_t, \quad  \forall l,k \\ &
0 \leq \tilde{r}_{l,k}, \tilde{m}_l \leq 1 ,  \quad \ \forall l,k. \label{eq:power_min_subprob1:binary_variables}
\end{align}
\end{subequations}
 For given, $r_{l,k}$, $m_l$, and $v_{l,k}$ values, \eqref{eq:power_min_opt_subprob1:problem} is in a convex form that can be solved by any convex programming solver. 

The second sub-problem will be solved only to find $v_{l,k}$, which can be described by
\begin{subequations} \label{eq:power_min_opt1:problem}
\begin{align}
 & \underset{\{ v_{l,k} \}}{\text{minimize}} \,   \lambda_3 \sum_{l=1}^{L} \sum_{k=1}^{K} (u_{l,k} - v^2_{l,k})^2 + \lambda_4 \sum_{l=1}^{L} \sum_{k=1}^{K} (\bar{\rho}_{l,k} - v_{l,k})^2.  
      \label{eq:power_min_opt1:objective} 
\end{align}
\end{subequations}
This problem serves two main purposes: (1) $\bar{\rho}^2_{l,k} \rightarrow u_{l,k}$ with the help of $v_{l,k}$, and (2) facilitate solving the first-problem jointly for $\tilde{M}_l$, $\bar{\rho}_{l,k}$. The solution for this problem is the positive real root of the following equation:
\begin{equation}
    4\lambda_3 v^3_{l,k} - (4\lambda_3 u_{l,k} - 2\lambda_4)v_{l,k} - 2\lambda_4\bar{\rho}_{l,k} = 0.
    \label{eq:subproblem2:solution}
\end{equation}

Finally, the third sub-problem is solved for the binary variables as 
\begin{subequations} \label{eq:subproblem3:problem}
\begin{align}
 & \underset{\{ r_{l,k}, m_l \}}{\text{minimize}} \,   c_2 \sum_{l=1}^{L} m_l  + c_3\sum_{l=1}^{L}  \sum_{k=1}^{K} r_{l,k} + c_4 \sum_{l=1}^{L} \tilde{M}_l \left( \sum_{k=1}^{K} r_{l,k} \right) \\ & +  \lambda_1 \sum_{l=1}^{L}  \sum_{k=1}^{K} (r_{l,k} - \tilde{r}_{l,k})^2 +  \lambda_2 \sum_{l=1}^{L} (m_{l} - \tilde{m}_{l})^2
      \label{eq:subproblem3:objective} \\ & \textrm{subject to} \nonumber \\ & 
r_{l,k}, m_l \in \{0,1\},  \quad \ \forall l,k. \label{eq:subproblem3:binary_variables}
\end{align}
\end{subequations}
Since only variables in this problem are the binary ones, the optimal solution of this sub-problem can be obtained by just checking the coefficients of these variables:
\begin{subequations} \label{eq:subproblem3:solution}
\begin{align}
& m_{l} = 0.5-0.5\cdot\mathrm{sign}\left(c_2 + \lambda_2 (1- 2\tilde{m}_l)\right)
 , \quad \forall l \\ & r_{l,k} = 0.5-0.5\cdot\mathrm{sign}\left( (c_3 + c_4 \tilde{M}_l) + \lambda_1 (1- 2\tilde{r}_{l,k}) \right) \quad \forall l,k. 
\end{align}
\end{subequations}
This way of separating variables simplifies the problem considerably by removing the integer optimization. 

The overall algorithm is described below. To guarantee a feasible start, the algorithm starts by activating all APs, and with random values of $v_{l,k}$. It first solves the sub-problem for continuous variables, \eqref{eq:power_min_opt_subprob1:problem}, then the second sub-problem for  $v_{l,k}$, \eqref{eq:subproblem2:solution},  and finally does the binary updates, \eqref{eq:subproblem3:solution}. Then we update the starting point and iterate over all sub-problems until the convergence. After convergence, we apply a post-processing procedure to efficiently obtain integer values of $M_l$ from continuous values of $\hat{M}_l=\tilde{M}_l+\tau_{\mathcal{S}_l}$. In the postprocessing, the algorithm first creates three different integer solution options by $M^{(1)}_l =  \lfloor \hat{M}_l \rfloor, M^{(2)}_l = \left[\hat{M}_l\right], M^{(3)}_l =  \lceil \hat{M}_l \rceil$, $\forall l$. Then, starting from the first option,  $M^{(1)}_l$, $m^{(1)}_l = \mathbb{I}(M^{(1)}_l)$, $\tilde{r}_{l,k}$ as an input, the post-processing algorithm  solves \eqref{eq:power_min_opt0:problem} with any convex programming solver.  If the solution is infeasible, the algorithm tries to solve the next option. If the solution is feasible,  the solution of the algorithm will be the final output.

\section{Simulation Results}

\begin{table}[tb!]
\caption{Simulation parameters}
    \vspace{-3mm}
\begin{tabular}{|l|l|l|l|}
\hline
$M^{\rm ac}$, $M^{\rm frh}$,  $M_{c}$ & 16, 64, 128 & $N_c$ & 4 \\ \hline
$f_s$, $B^{\rm ac}$, $B^{\rm frh}$ & $30.72$, $20$, $100$\,MHz   &  $T_s$ & $71.4\,\mu$s \\ \hline
$P_t$, $P_f$, pilot pow.  & $1$, $5$, $0.1$\,W  & $P_{\mathrm{fixed}}$  & $120$\,W  \\ \hline
 $\tau_c$, $\tau_p$ &  $192$, $8$ & $\sigma_{\mathrm{cool}}$  & $0.9$  \\ \hline
$C_{\mathrm{GPP}}^{\max }$, $C_{\mathrm{AP, max}}$ & $180$ GOPS &  $P_{\mathrm{st}}$ & $6.8$\,W \\ \hline
$\Delta^{\mathrm{proc}}_{\mathrm{AP}}, \Delta^{\mathrm{proc}}_{\mathrm{GPP}}$ & $74$\,W & $P^{\mathrm{proc}}_{0}$ & $20.8$\,W \\ \hline
$N_{\mathrm{used}}$, $N_{\mathrm{bits}}$ &  1024, 12 & $P_{\mathrm{comp}}$ & $20.8$\,W \\ \hline
\end{tabular}
\label{tab:simulation_params}
\end{table}
We consider a square area of size  $1 \times 1~\text{km}^2$ with grid type AP deployment, where the cloud processor is located in the center of the area. There are $L=16$ APs. We consider $2.5$\,GHz and $28$\,GHz carrier frequency for the access and fronthaul links, respectively. The remaining simulation parameters are given in Table \ref{tab:simulation_params}. The UEs are distributed uniformly in the considered area. We run $200$ random simulations and take the average of the performance results. 
\begin{figure}[tb]
    \centering
    \includegraphics[width=\linewidth]{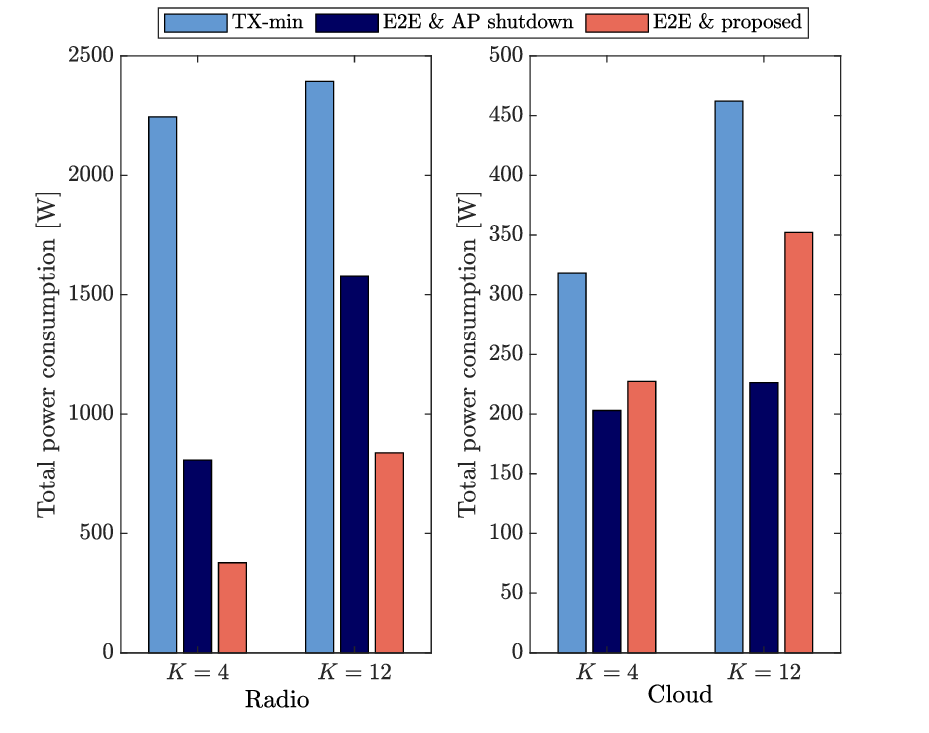}
      \vspace{-8mm}
    \caption{Power consumption performance of different algorithms considering wireless fronthaul. TX-min considers only minimizing transmit power from APs to UEs, while E2E \& AP shutdown minimizes end-to-end power consumption by jointly deciding AP activation and power allocation. The proposed algorithm minimizes end-to-end power consumption by also optimizing the utilized number of antennas at each AP.}
    \label{fig:all_algs_comp}
    \vspace{-6mm}
\end{figure}

Fig. \ref{fig:all_algs_comp} compares the power consumption of the proposed algorithm and other baselines considering wireless fronthaul. Radio and cloud demonstrate the power consumption at the  APs and cloud due to the transmit power and the processing as described in Section II.C.  TX-min is a baseline, where we only aim to minimize the total transmit power of the APs in the access channel. This is a scheme that is widely utilized in the literature \cite{cell-free-book}. E2E \& AP shutdown is the baseline algorithm and inspired by \cite{ozlem_jsac} considering wireless fronthaul constraints. We can observe that power consumption on the radio is much higher than the cloud since the precoding and DFT operations are done at the AP side, requiring the highest power.  An obvious observation is that the total power consumption of both AP and antenna activation schemes are much lower than the TX-min algorithm, demonstrating the importance of considering end-to-end energy consumption in the resource allocation. While the proposed scheme requires higher power consumption in the cloud side, considering the total power consumption, it reduces the power consumption more than $40\%$ compared to the AP shutdown algorithm, demonstrating the advantage of the proposed scheme.

      \vspace{-3mm}

\section{Conclusion}
In this work, we have considered a wireless fronthaul cell-free massive MIMO network to reduce the deployment cost by removing the deployment of fiber cables. In order to allocate the power and time resources in the fronthaul, we have proposed a mix of TDMA/SDMA scheme, and optimal AP grouping algorithm where APs are grouped to minimize the interference within a group.  We have proposed a joint antenna activation and power allocation algorithm that provides load balancing not only in the access channel but also in the fronthaul channel to minimize end-to-end power consumption in the network. Our numerical results demonstrate that the proposed algorithm can  reduce the power consumption by  $40\%$ compared to the AP shutdown baseline algorithm. As future work, the proposed methodology can be applied considering centralized precoding operation, and lower layer functional splits to further reduce the power consumption of the network. 

\bibliographystyle{IEEEtran}
\bibliography{IEEEabrv,refs}

\end{document}